# Modelling PM10 Crisis Peaks Using Multi-Agent based Simulation: Application to Annaba City, North-East Algeria

Sabri Ghazi[1], Julie Dugdale[2,3] and Tarek Khadir[1]

[1] Laboratoire de Gestion Electronique de Documents, Department
of Computer Science, University Badji Mokhtar, PO-Box 12, 23000,
Annaba, Algeria.
[2] University Grenoble Alps. LIG. France.
[3] University of Agder, Norway.
`Ghazi@labged.net;Khadir@labged.net;Julie.Dugdale@imag.fr`

**Abstract.** The paper describes a MAS (multi-agent system) simulation approach for controlling PM10 (Particulate Matter) crisis peaks. A dispersion model is used with an Artificial Neural Network (ANN) to predict the PM10 concentration level. The dispersion and ANN models are integrated into a MAS system. PM10 source controllers are modelled as software agents. The MAS is composed of agents that cooperate with each other for reducing their emissions and control the air pollution peaks. Different control strategies are simulated and compared using data from Annaba (North-East Algeria). The simulator helps to compare and assess the efficiency of policies to control peaks in PM10.

**Keywords:** Multi-Agent System, Multi-Agent Based Simulation, Agent-Based Modelling, Environmental Modelling, Air Pollution, Air Quality, PM10.

## 1 Introduction

PM10 (Particulate Matter with an aerodynamic diameter of 10 micrometers) is a complex mixture of polluted air, including organic and inorganic particles. The toxic build-up of PM10 may be considered as a slow onset crisis and is a major health threat in many cities throughout the world [1]. This may result in large social and economic costs. Many factors influence the concentration of PM10. Some are natural such as the local climatic and topographic conditions, while others are manmade, such as vehicle emissions and the result of industrial activities. A crisis peak occurs when the concentration of air pollutant exceeds pre-defined levels, thus concentration levels need to be accurately predicted and controlled. Simulation and decision support tools are valuable for decision-makers in order to assess the efficiency of their policies in the management of pollution during peaks periods.

Many air pollution modelling approaches have been proposed such as: mathematical emission models ([2], [3],), linear models, ANN (Artificial Neural Networks) models ([4]) and hybrid models [5]. Most of them only address the physical and chemical aspects of air pollution (concentration and dispersion) and do not consider human decision-making factors. Air pollution is, by nature, distributed and results from the complex interaction of many actors. Anthropogenic activities (road traffic, industrial and agricultural activities) are among the major sources of air pollution. Therefore, it is essential to include human decision-making factors in the modelling of air pollution. A multi-agent system (MAS) allows us to model the behaviours of human actors sharing the exploitation of environmental resources [6] and it is an appropriate method for simulating pollution related issues. [7] used a MAS approach to investigate the air pollution emission resulting from road activities by using a traffic flow MAS simulation linked to an emission calculation. A methodology for building an Environmental Information System (EIS) based on a MAS is presented in [8]. In that work an agent, which represents a human-being or a group of humans-beings, uses case based reasoning to make decisions. The methodology was used to develop a system for air quality reporting. [9] presents ECROUB, a MAS for managing the quality of an urban microclimate. Using physical models, the system was able to generate information about the climate in a very small geographic zone. The system shows that a hybrid approach (MAS and physical models) can be productively used for studying urban areas.

[10] describes a MAS system designed for monitoring air quality in Athens, Greece. It is composed of a set of software agents, controlling a network of sensors installed in different positions of an urban region. Agents verify the data measured by sensors. The system uses a prediction given by an ANN model. Real data about ozone concentration and meteorological data was used to feed the simulation system. In [11] a MAS is used to model air pollution in an urban area. The environment is represented by a two-dimensional grid. The purpose of the simulation is to find the dispersal of air pollution on the grid. Each cell of the grid has a value of pollutant concentration. Neighbours with a close pollution rate (according to an initially set threshold) form a cluster. The pollution sources are represented by homogeneous agents that emit pollution in their areas (polluters). Each agent pollutes according to its emission rate. As the simulation runs, clusters are formed with different values of pollution concentration. Although similar to our approach, this model does not include meteorological parameters, does not address a specific pollutant types, and does not use real data.

We aim to model and simulate the possible cooperation between pollution source controllers and investigate the impact of cooperation on PM10 concentration. The simulator focuses on PM10 crisis peaks and the regulations that should be adopted in order to manage and reduce its effect. To investigate these questions we present a MAS approach for simulating PM10 concentrations. The feasibility of our approach is demonstrated by a scenario using data from Annaba, a Mediterranean city in the northeast of Algeria.

The paper is organized as follows: Section (2) describes the architecture of the simulator and the representation of the environment. We also explain the dispersion and prediction models and their integration, and define the cooperation strategies. Section (3) presents a description of the simulation scenario using the data from Annaba city. The results of the simulation are presented and discussed in section (4). We end the paper by a conclusion and possible future work.

## 2  Model Approach and Architecture

Our simulation approach models the agents' actions that affect the emission rate of the sources they control. The dispersion algorithm is then used to compute how PM10 spreads in the environment; the aggregated value of pollutant concentration is used together with climatic parameters to forecast the air pollution concentration *k* hours ahead. According to these forecasts, agents are rewarded or penalised using a regulation formula that takes into account how the agent has contributed to peak concentration. Agents then adapt their strategies to earn more reward and/or reduce penalties.

### 2.1   The spatial and temporal scale of the simulation model

The simulator uses a discrete representation of time where each simulation step represents 2 hours of real time. The environment is modelled as a set of *3D* boxes, each one represents one KM$^3$, every box is localised at *gp(x,y,z)* and has an attribute representing the concentration of PM10.

### 2.2   Dispersion and prediction models

The dispersion model describes how the pollutant will spread in the air. It is calculated according to the distance from the point source, the wind speed and the emission rate. We used a GPD (Gaussian Plum Dispersion) model (1), which is frequently used in atmospheric dispersion [12].

$$C(x,y,z,H) = \frac{er_{i,t}}{2\pi U_t \, \sigma_y \sigma_z} * e^{-\frac{y^2}{2*\sigma_y^2}} * \left[ (e^{-(\frac{(z-H)^2}{2\sigma_z^2})}) + (e^{-(\frac{(z+H)^2}{2\sigma_z^2})}) \right] \quad (1)$$

The concentration of PM10 is calculated according to: $er_{i,t}$: the emission rate in kilograms per hour of the source *i* at time step *t*, and $U_t$: the wind speed in metres per second at time step *t*, $\sigma_y \sigma_z$: the standard deviation of the concentration distributions in vertical direction crosswind.
The level of pollution resulting from each source is aggregated and the average per box is computed. Then the dispersal value of the PM10 is passed to an ANN prediction model. The ANN prediction model is designed to give a forecast of the air pollutant and also the overall air quality. This includes an uncertainty aspect caused by the weather conditions. The ANN predictor uses the aggregated air pollution concentration value from the dispersal model of each source and the four climatic parameters: wind speed, humidity, temperature and rainfall. These parameters greatly influence the pollutant concentration.

### 2.3    Decision-making mechanism

Based on its internal state (the value of its internal attributes) and the state of the environment (values of variables representing the environment), an agent has to choose which action to perform among all of its possible actions in order to reach its goals. This process is called decision-making. Our system supports two cooperation strategies (centralized and evolutionary game) each one defines a decision-making mechanism. The centralized strategy (CS) is based on defining a central agent that represents the air pollution control agency. The central agent makes decisions according to the current air pollution level. The second strategy is based on an evolutionary game, where agents are rewarded and penalized according to the pollution level; making decisions according to their rewards. In our system, the cooperation strategy is defined as part of the simulation parameters.

**Centralized Strategy (CS):**
The task of maintaining the air quality is assigned to an agent, which represents the air pollution control agency. It uses the prediction about air quality and pollutant levels, and accordingly sends a reduce emission message to the emission agents. Then it will recheck the air quality. It will continue doing this until the end of the peak period. As in the real world situation, the central agent has sufficient authority to ensure that the emission source controllers execute its orders. Agents communicate their emission rate at each simulation step. We assume that agents are rational and are environmentally responsible, favouring air quality improvement over their own interests and communicating to the central agent their exact emission rate.

**Evolutionary Game Cooperating Strategy**
In the EG strategy, every agent has its own goals (earning more rewards and keeping its emission rate as high as possible) and shares a global goal of maintaining air quality with other agents. An agent participates with other agents in the game, its own goal is to maximise its reward earned from the game. We adopted the approach of [13], where agents keep traces of their *K* previous steps (actions, rewards and its neighbours' rewards). At each time step *t* the agent computes its weighted payoff according to (4) and updates its probability of increasing or decreasing its emission rate, respectively according to (5) and (6).

$$WP_t = \sum_{i=1}^{k} w_i * M_i \qquad (2)$$

Where: $w_i$ is the weighting parameter where $\sum_{i=1}^{k-1} w_i = 1$ and $\forall i,j \ (i < j \rightarrow w_i > w_j)$, $M_i$ is the *i-th* payoff, *i=1* means the payoff earned this step.

$$\begin{cases} Pc_i(t+1) = Pc_i(t) + 1 - Pc_i(t) * \alpha, if \ S_i = 0 \ and \ WP_t > 0 \\ Pc_i(t+1) = (1-\alpha) * Pc_i(t), \qquad if \ S_i = 0 \ and \ WP_t \leq 0 \end{cases} \qquad (3)$$

$$\begin{cases} Q_i(t+1) = Q_i(t) + 1 - Q_i(t) * \alpha, & if \ S_i = 1 \ and \ WP_t > 0 \\ Q_i(t+1) = (1-\alpha) * Q(t), & if \ S_i = 1 \ and \ WP_t \leq 0 \end{cases} \quad (4)$$

Where: $Pc_i$ and $Q_i$ are respectively the probability to decrease (*S=0*) and increase (S=1) the emission for the agent *i*, *α* is the learning rate, *S* is the strategy played at time *t*. Agents are influenced by their neighbours at each time step; the average reward of the neighbours is calculated according to (9).

$$nP_{i,j}(t) = (\sum_{j=1}^{R} C_j)/R \quad (5)$$

Where *Cj* is the payoff of the neighbour *j*, and *R* is the number of neighbours for the *i-th* agent. The average of the *K* last *nP* is noted *avgNP*. The agent then uses the probabilities *Pc, Pd* and the average reward of its neighbours to choose an action according to the algorithm below:

```
Algorithm    Choose Action Emission Agent
while t < MaxTemps do
  if (lastChoices[0] == 0) then
    if ((RPwt < pf_AVG) and (P < Q) and (Q > Ru)) then
      lastChoices[0] = 1;
      sourcesInfo.resumeEmission();
    else
      lastChoices[0] = 0;
      sourcesInfo.reduceEmission();
    end if
  end if
  if (lastChoices[0] == 1) then
    if ((RPwt < pf_AVG) and (Q < P) and (P > Ru)) then
      lastChoices[0] = 0;
      sourcesInfo.reduceEmission();
    else
      lastChoices[0] = 1;
      sourcesInfo.resumeEmission();
    end if
  end if
end while
```

**Fig.1.** Algorithm for choosing an action using the EG strategy

When a crisis peak occurs, the system uses an agent's emission rate to calculate how much the agent has contributed to the current level of the pollution. The agent is penalised according to its level of participation.

## 3  Simulation scenarios using data from the Annaba region

The dataset used in this work covers the period 2003-2004 on a continuous basis of 24 hours. Air pollutants, including PM10, are continuously monitored. The dataset also includes four meteorological parameters: Wind Speed (WS), Temperature (T), relative Humidity (H) and rainfall.

A simulation scenario for the region of Annaba was defined to include a 100 sources of PM10 with the maximum emission rate for each source being 2000 gram/hour. The goal level for pollutant concentration is fixed according to the air quality standards and must be bellow 70 microgram per cubic meter. The initial values (at *t=0*) for the concentration of pollutant and the climatic parameters are fixed according to the dataset. For the case of EG strategies we fixed the initial proportion of cooperating agents (agents choosing to decrease emission) to *0.5*, this means that *50%* of the agents decrease their emission at *t=0*. This proportion will change during the simulation according to the game outcome. The prediction is 4 hours in advance, the same as the simulation step. Each source emits pollutants according to its emission rate, which cannot be higher than the maximum level defined in the simulation scenario. The position of sources is randomly generated.

## 4      Results and discussion

The scenario was run by choosing at each time a cooperation strategy: (with Penalties), (No Penalties), CS (Centralized Strategy) and NC (No-Cooperation). The last one is included for comparison purposes. Due to space limitations only the most significant results are presented.  When executed 80 times the simulation showed that the CS gives similar results at each run, the same thing was also found for NC. Using the EG strategies, the simulations show slight differences between runs especially in the proportion of cooperating agents. These changes are due to the random values used in the initialisation of some variables (neighbours rewards, first chosen action, weights, *k* last actions and rewards).

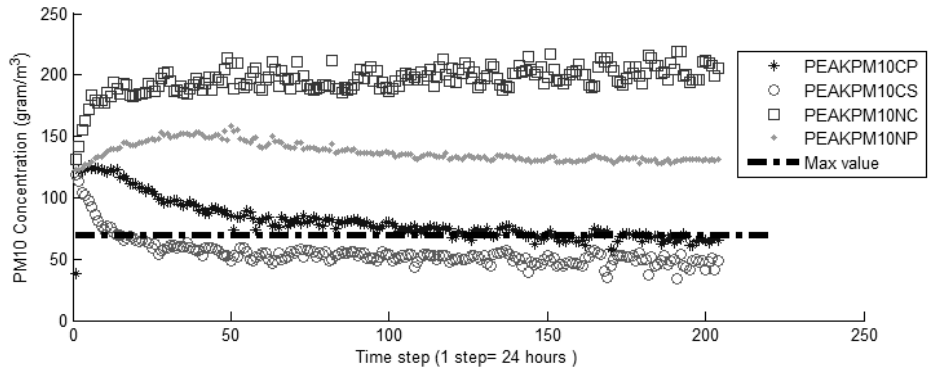

**Fig.2.** Peak of 24 hours of PM10 for the four tested controlling strategies compared with the no-cooperation strategy.

Figure 2 shows the concentration of PM10 for each strategy during the simulation; each point is the peak level in 24 hours. The CS strategy (PICPM10CS) performs the best, taking less time to bring the pollution level under control and keeping it below the goal level defined in the simulation scenario. The EG strategy (PICPM10CP) performs reasonably well, but it takes slightly longer than the CS strategy to bring the pollution level down. Nevertheless it manages to keep the pollution level near to the goal. The penalising regulations have a big effect on the PM10 level. As illustrated, the PICPM10CP (with penalties) controls the pollution and performs better than the non-penalising strategy (PICPM10NP). The NC strategy (No-Cooperation) is presented in order to show the impact of cooperation on the PM10 level. The CS gives the best performance since the pollution concentration rapidly decreases. When cooperation is not used (PICPM10NC), agents act selfishly and do not care about the pollution. As the agents reach their maximum emission rate, we can observe an oscillation that is caused by the climatic conditions. Consequentely, the pollutant level reaches alarming values and many peaks periods occur.

## 5      Conclusions

Computer simulations are a valuable tool for helping in the management of pollution related crisis. In our study we showed how a multi-agent based simulation approach could be successfully combined with classic modelling tools in order to model air pollution peaks. The simulation helps to investigate different controlling strategies by measuring how each strategy performs in managing the crisis peaks of PM10.

The regulation rules are computed according to how much the agent participates to the level of pollution; it is clear that this regulation has a big influence in controlling pollution. As shown in the simulation results, cooperation helps to reduce the pollution level and it also affects the evolution of the pollutant. This is especially noticeable during the peak periods where climatic conditions cause the pollutants to stagnate.

The current version of the system only models point emission sources of PM10. In future versions we aim to include continuous sources, such as roads, and to predict other pollutants. The simulator may also be enhanced by including a GIS interface. In addition, exploring other cooperation strategies are also among our future plans.

## 6      Acknowledgements

This work was funded by the Algerian Ministry of Higher Education and Scientific Research, PNE 2014/2015 Program.